\documentstyle[hip-artc]{article}  
\newcommand{\dlt}{\Delta}
\newcommand{\beq}{\begin{equation}}
\newcommand{\eeq}[1]{\label{#1} \end{equation}}
\newcommand{\beqar}{\begin{eqnarray}}
\newcommand{\eeqar}[1]{\label{#1} \end{eqnarray}}
\newcommand{\g}{\gamma}
\volnumber{0}  \edyear{0000}  \frompage{000} \topage{000}                
\recrevdate{Version 1.03     25.02.99} 		 
\def\rightmark{Kinetic freeze out} 
\def\leftmark{V.K. Magas, et al.}
 
\title{Kinetic freeze out models} 
\authors{
{\twerm 
V.K. Magas$^1$, 
Cs. Anderlik$^{1,2}$, 
L.P. Csernai$^{1,2,3}$, 
F. Grassi$^4$,\\
W. Greiner$^2$,
Y.~Hama$^4$, 
T.~Kodama$^5$,
Zs.I. L\'az\'ar$^{1,2,6}$ 
and 
H. St\"ocker$^2$%
}\\[2.812mm]
{\normalsize
\hspace*{-8pt}$^1$ Section for Theoretical Physics, Department of Physics\\
University of Bergen, Allegaten 55, 5007 Bergen, Norway\\[0.2ex]
\hspace*{-8pt}$^2$ Institut f\"ur Theoretische Physik, Universit\"at Frankfurt\\
Robert-Mayer-Str. 8-10, D-60054 Frankfurt am Main, Germany\\[0.2ex]
\hspace*{-8pt}$^3$ KFKI Research Institute for Particle and Nuclear Physics\\
P.O.Box 49,  1525 Budapest, Hungary\\[0.2ex] 
\hspace*{-8pt}$^4$Instituto de F\'{\i}sica, Universidade de Sao Paulo\\
CP 66318, 05389-970 S\~ao Paulo-SP, Brazil\\[0.2ex]
\hspace*{-8pt}$^5$Inst. de F\'{\i}sica, Universidade Federal do Rio de Janeiro\\
CP 68528, 21945-970 Rio de Janeiro-RJ, Brazil\\[0.2ex] 
\hspace*{-8pt}$^6$Department of Physics,
Babe\c{s}-Bolyai University\\Str. M. Kog\u{a}lniceanu nr. 1, 3400
Cluj-Napoca, Romania }}
 
\abstract{
Freeze out of particles across a space-time hypersurface is discussed in 
kinetic models. The calculation of final momentum distribution of emitted 
particles is described for freeze out surfaces, with spacelike normals. 
The resulting non-equilibrium distribution does not resemble, the previously
proposed, cut J\"uttner distribution, and shows non-exponential $p_t$-spectra
similar to the ones observed in experiments.
} 

 
\begin{document}
 
\maketitle

\section{Introduction}

Continuum and fluid dynamical models, due to their simplicity are very
popular in heavy ion physics, because they connect directly collective
macroscopic matter properties, like the Equation of State (EoS) or transport
properties, to measurables.

In contrast to kinetic, Monte Carlo models, in continuum models the
evaluation of measurables is a problem in itself.  Particles, which leave
the system and reach the detectors, can be taken into account via source
(drain) terms in the 4-dimensional space-time based on kinetic
considerations, or in a more simplified way via freeze out (FO) or final
break-up schemes, where the frozen out particles are formed on a
\mbox{3-dimensional} hypersurface in space-time. Such freeze out descriptions 
are important ingredients of evaluations of two-particle correlation data, 
\mbox{transverse-,} \mbox{longitudinal-,} radial-, and cylindrical-
flow analyses, transverse momentum and transverse mass spectra and many other
observables.

The general theory of discontinuities in relativistic flow was not worked out
for a long time, and the 1948 work of A.  Taub \cite{Ta48} discussed
discontinuities across propagating hypersurfaces only (which have a
spacelike unit normal vector $d\hat{\sigma}^\mu$,
 $d\hat{\sigma}^\mu d\hat{\sigma}_\mu = -1$).
Events happening on a propagating, (2 dimensional) surface belong to this
category.

Another type of change in a continuum is an overall sudden change in a finite
volume. This is represented by a hypersurface with a timelike normal
($d\hat{\sigma}^\mu d\hat{\sigma}_\mu = + 1$). If one
applies Taub's formalism to freeze out surfaces with timelike normal
vectors, one gets a usual Taub adiabat, but the equation of the Rayleigh line
will yield imaginary values for the particle current across the
front. Only
in 1987 Taub's approach was generalized to both types of surfaces \cite{Cs87},
making it possible to take into account conservation laws exactly across any
surface of discontinuity in relativistic flow.  This approach also eliminates
the imaginary particle currents arising from the equation of the Rayleigh
line. When the EoS is different on the two sides of the FO front these
conservation laws yield changing temperature, density, flow velocity across
the front. Based on conservation laws and on simple kinetic considerations
the idealized surface freeze out models  were completely worked out in refs.
\cite{CF74,Bu96,AC98,ALC98}.

\section{Conservation laws across idealized freeze out fronts} 
\label{three}

The FO hypersurface is an idealization of a layer of
finite thickness (of the order of a mean free path or collision time) where
the frozen-out particles are formed and the interactions in the matter become
negligible. The dynamics of this layer can be described in different kinetic
models as Monte Carlo models~\cite{BB95,BB97} or four-volume emission
models \cite{Barz82,GH95,GH96,GH97,He97}. The zero thickness limit of such a
layer is the idealized FO surface. Kinetic models for hadronic degrees of
freedom indicate that such an idealization is meaningful only for collisions
of massive heavy ions like Au+Au or Pb+Pb \cite{BB95,BB97}. If we include
quark-gluon plasma in our reaction model with rapid final hadronization which
coincides with freeze out \cite{CC94,CM95}, the applicability of idealized
surface freeze out description becomes even better.

In general the energy - momentum tensor changes discontinuously across this
idealized hypersurface.  If the flow is not orthogonal to this surface the
four-vector of the flow velocity will also change across this
surface \cite{Cs87,CC94,CG88}. The FO discontinuity has a timelike normal in
most cases, and so the method for the description of timelike detonations
and deflagrations~\cite{Cs87} should be used (see also
\cite{CC94,CG88,GC86,LP90,Go90}).

If $d\sigma^\mu$ is the normal vector of an element of the FO hypersurface
 ($d\hat{\sigma}^\mu$ is a unit vector parallel to $d\sigma^\mu$),
the observable triple differential cross section is obtained by integrating
the FO current over the whole FO hypersurface. This is described by the
Cooper-Frye formula \cite{CF74,AC98}:
\beq
E\frac{dN}{d^3p} = \int f_{FO}(x,p;T,n,u^\nu) \ p^\mu d\sigma_\mu\ ,
\eeq{e-cf}
where $f_{FO}(x,p;T,n,u^\nu)$ is the post FO phase space distribution of
frozen out particles which is not known from the fluid dynamical model.  When
we apply this formula we have to ensure \cite{AC98} that: (i) the particles
are really leaving the surface outwards:
\beq
f_{FO}(x,p;T,n,u^\nu) =
\Theta(p^\mu\ d\hat{\sigma}_\mu) 
f_{FO}(x,p;T,n,u^\nu)  \,,
\eeq{fthef}
and  (ii) the conservation laws and entropy condition are satisfied%
~\cite{Ta48,Cs87}:
\beq
[N^\mu\ d\hat{\sigma}_\mu] = 0 \,,  \ \ \ \ \ \ 
[T^{\mu\nu}\ d\hat{\sigma}_\mu] = 0 \ \ \ \ \  {\rm and} \ \ \ \ \ \
[S^\mu\ d\hat{\sigma}_\mu] \ge 0 \,,  \ \ \ \ \ \ 
\eeq{efo2}
where $[A]\equiv A - A_0$, and the pre FO baryon and entropy currents and
energy-momentum tensor are $N_0^\mu$, $S_0^\mu$ and $T_0^{\mu\nu}$, while
the post freeze out quantities are, $N^\mu$ $S^\mu$ and $T^{\mu\nu}$, In
numerical calculations the local freeze out surface can be determined most
accurately via self-consistent iteration \cite{Bu96,NL97}. This fixes the
parameters of our post FO momentum distribution, $f_{FO}(x,p;T,n,u^\nu)$ if
the shape of the distribution is known.

A simple approach is to assume that $f_{FO}$ is a cut J\"uttner distribution
\cite{Bu96,AC98} if we have a FO hypersurface with spacelike normal.  It was
shown that one can formally solve the freeze out problem with this
ansatz \cite{AC98}.

\section{Freeze out distribution from kinetic theory}
\label{six}

At the same time in a kinetic gas model it was studied how we can obtain a
non-equilibrium post FO distribution across a FO hypersurface with a 
spacelike normal \cite{ALC98}. The result indicated that actually the cut J\"uttner
distribution is not a realistic choice. Apart of the non-physical shape by
the sharp momentum cut, the simplified model which recovered the cut
J\"uttner shape led to other nonphysical consequences also: The freeze out
was not complete in the model even for an asymptotically large thickness of
the FO layer.  Thus, the simple kinetic model \cite{ALC98} had two
unsatisfactory features: (i) it did not achieve complete freeze out, and
(ii) it resulted in exponentially weakening freeze out with a FO layer
thickness which tends to infinity. A solution to the first problem has also
been suggested.

Here we present solutions to both problems by making additional improvements
of the model used in \cite{ALC98}. The
 two extended models are still very superficial and should not be
considered as final and physically realistic solutions of the freeze out
problem. For transparency we perform and demonstrate the modifications
separately, although both of these could be included in a combined model.
This way we can see which improvement (consideration of which physical
process) cures problems.

Let us assume an infinitely long tube with its left half ($x<0$) filled with
nuclear matter and in the right vacuum is maintained. We can remove the
dividing wall at $t=0$, and then the matter will expand into the vacuum. By
continuously removing particles at the right end of the tube and supplying
particles on the left end, we can establish a stationary flow in the tube,
where the particles will gradually freeze out in an exponential rarefaction
wave propagating to the left in the matter.  We can move with this 
rarefaction wave or front, so
that we describe it from the reference frame of the front (RFF), 
where the rarefaction is stationary.

In this frame, we have a stationary supply of equilibrated matter from the
left, and a stationary rarefaction front on the right, $x>0$.  We can
describe the freeze out kinetics on the r.h.s. of the tube assuming that we
have two components of our momentum distribution \cite{GH95,GH96,GH97},
$f_{free}(x,\vec{p})$ and $f_{int}(x,\vec{p})$. However, we only assume that
at $x=0$, $f_{free}$ vanishes exactly and $f_{int}$ is an ideal J\"uttner
distribution (supplied by the inflow of equilibrated matter), while $f_{int}$
gradually disappears and $f_{free}$ gradually builds up as $x$ tends to
infinity.

Let us recall \cite{ALC98} the most simple kinetic model describing the
evolution of such a system. Starting from a fully equilibrated J\"uttner
distribution the two components of the momentum distribution develop
according to the coupled differential equations:
\beqar
\partial_x f_{int}(x,\vec{p})   dx &=& - \Theta(p^\mu d\hat{\sigma}_\mu) 
                                   \frac{\cos \theta_{\vec{p}} }{\lambda}
           f_{int}(x,\vec{p})   dx,\nonumber
 \   \\ 
\partial_x f_{free}(x,\vec{p})  dx &=& + \Theta(p^\mu d\hat{\sigma}_\mu) 
                                   \frac{\cos \theta_{\vec{p}} }{\lambda}
           f_{int}(x,\vec{p})   dx.
\eeqar{kin-1}

Here $\cos \theta_{\vec{P}} = \frac{p_{x}}{p}$ in the RFF frame.  The
interacting component, $f_{int}$, will deviate from the J\"uttner shape and
the solution will take the form:
\beq
f_{int}(x,\vec{p}) =  f_{Juttner}(x=0,\vec{p}) 
\exp \left[ - \Theta(p^\mu d\hat{\sigma}_\mu) 
              \frac{\cos \theta_{\vec{p}} }{\lambda}  x \right].
\eeq{sol-11}
This solution is depleted in the forward $\vec{p}$-direction, particularly
along the $x$-axis.  Inserting it into the second differential equation
above, leads to the freeze out solution:
\beqar
f_{free}(x,\vec{p}) =  f_{Juttner}(x=0,\vec{p}) 
\left\{1-\exp\left[-\Theta(p^\mu d\hat{\sigma}_\mu)
\frac{\cos\theta_{\vec{p}}}{\lambda}x\right]\right\} = \nonumber \\ & & \\
f_{Juttner}(x=0,\vec{p})  - f_{int}(x,\vec{p}) \,. \nonumber
\eeqar{sol-12}
At $x \longrightarrow \infty$ this distribution will tend to the cut
J\"uttner distribution introduced in the previous section.  The remainder of
the original J\"uttner distribution survives as $f_{int}$, even if $x
\longrightarrow \infty$. In this model the particle density does not change 
with $x$, barely particles moving faster than the freeze out front (i.e.
\mbox{$p^\mu d\hat{\sigma}_\mu  > 0$)} are transferred gradually from component 
$f_{int}$ to component $f_{free}$.  This is a highly unrealistic model,
indicating that rescattering and re-thermalization should be taken into
account in $f_{int}$. This would allow particle transfer from the "negative
momentum part" (i.e. $p^\mu d\hat{\sigma}_\mu < 0$) of $f_{int}$ to $f_{free}$,
which is not possible otherwise.

\section{Freeze out distribution with rescattering}
\label{seven}

The assumption that the interacting part of the distribution remains the
distorted (after some drain) J\"uttner distribution, is of course highly
unrealistic.  As suggested in ref. \cite{ALC98} rescattering within this
component will lead to re-thermalization and re-equilibration of this
component. Thus the evolution of the component, $f_{int}$ is determined by
drain terms as much as the re-equilibration.  Here, we repeat the
introduction of the model with rescattering, and present it more generally
for both positive and negative flow parameter velocities. This turned out
necessary because during the FO process the interacting component will
acquire negative velocity parameters, even with large positive initial
velocities. We also introduce the $m=0$ limit of the suggested solution which
enables us to present a transparent and not completely numerical solution.

If we include the collision terms explicitly into the transport equations
(\ref{kin-1}) in general case leads to a combined set of integro-differential
equations. We can, however, take advantage of the  relaxation 
time approximation to simplify the description of the dynamics.

Then 
the two components of the momentum distribution develop according to the
coupled differential equations:
\beq
\begin{array}{rll}
\partial_x f_{int}(x,\vec{p}) dx =& - \Theta(p^\mu d\hat{\sigma}_\mu) 
                                   \frac{\cos \theta_{\vec{p}} }{\lambda}
           f_{int}(x,\vec{p})   dx+
 \\ & & \\
                                  
           &+\left[ f_{eq}(x,\vec{p}) -  f_{int}(x,\vec{p})\right]
           \frac{1}{\lambda'} dx, 
\end{array}
\eeq{kin-2}
\beq
\begin{array}{rll}
\partial_x f_{free}(x,\vec{p}) dx =& + \Theta(p^\mu d\hat{\sigma}_\mu) 
                                   \frac{\cos \theta_{\vec{p}} }{\lambda}
           f_{int}(x,\vec{p})  dx.
\end{array}
\eeq{kin-3}

The interacting component of the momentum distribution, described by \linebreak
eq. (\ref{kin-2}), shows the tendency to approach an equilibrated
distribution with a relaxation length  $\lambda' $.  Of course
due to the energy, momentum and conserved particle drain, this distribution,
$f_{eq}(x,\vec{p})$ is not the same as the initial J\"uttner distribution,
but its parameters, $n_{eq}(x)$, $T_{eq}(x)$ and $u^\mu_{eq}(x)$,
change as required by the conservation laws.


\begin{figure}[hp]
	\insertplot{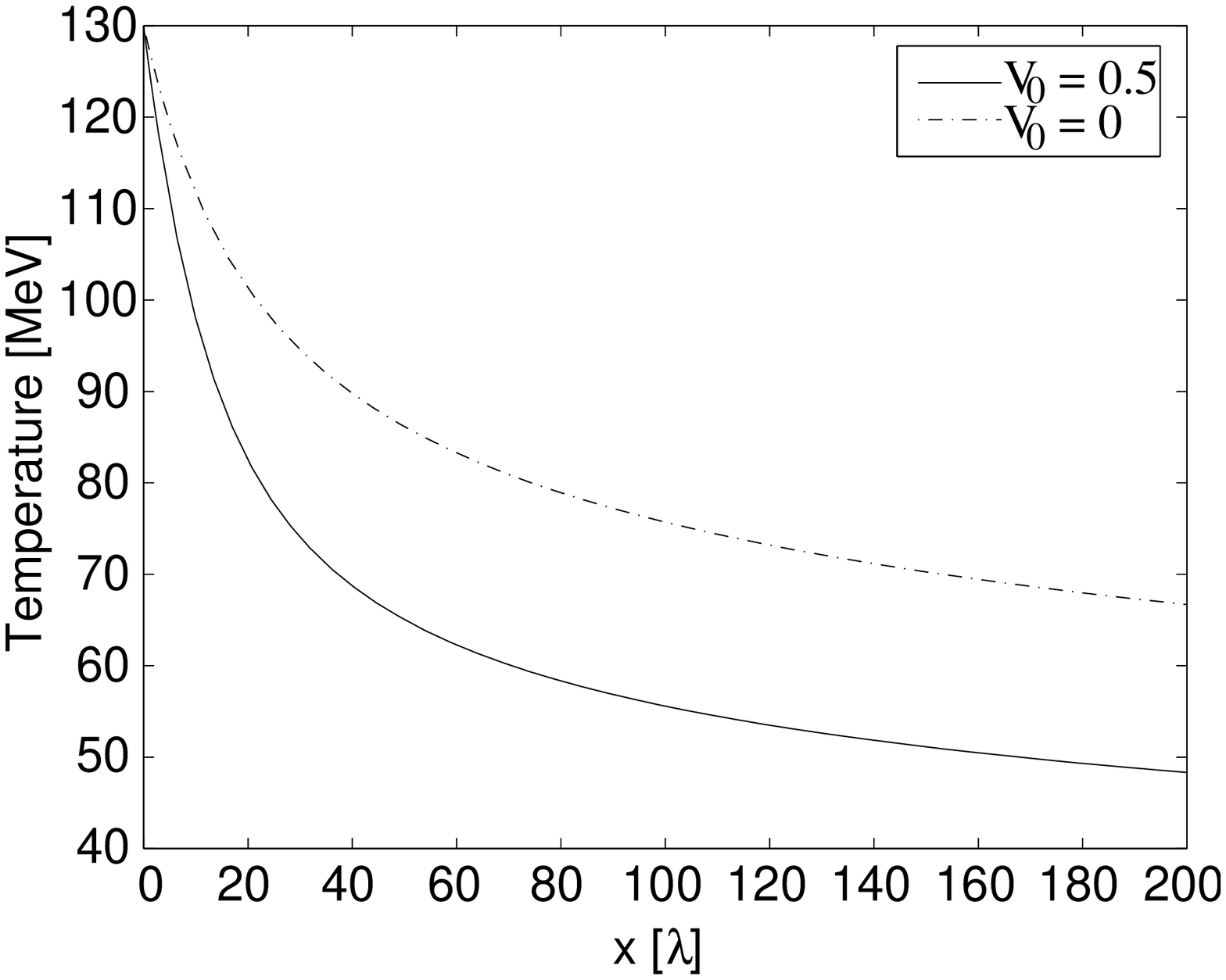}
\vspace*{-0.2cm}
	\insertplot{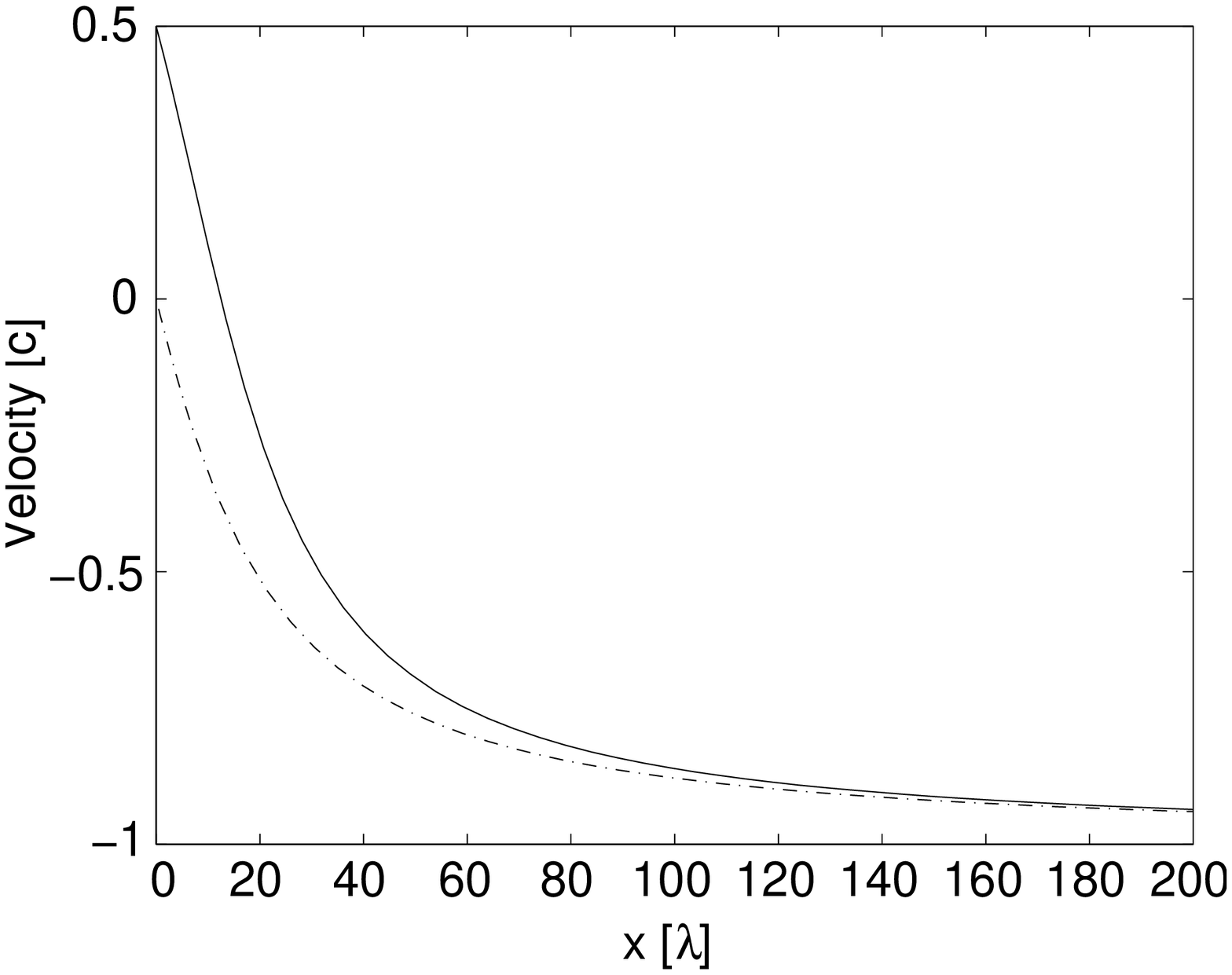}
\vspace*{-0.6cm}
\caption[]{
 The temperature and 
flow velocity of the interacting component 
in the rest frame
of the freeze out front (RFF)
for baryonfree and massless gas,
calculated according to the 
coupled set of equations (\ref{simmod}).
We assume an initial temperature of
\mbox{$T_0=130$ MeV} to fit the
situation in refs. \cite{CC94,CM95}.}

\label{fig:2}
\end{figure}

{\em Conservation Laws.} \ \ 
In this case the change of the conserved quantities 
caused by the particle transfer from component $int$ to component $free$
can be obtained in terms of the distribution functions as:
\beq
d N_i^\mu = 
      -\frac{dx}{\lambda}\int\frac{d^3p}{p_0} p^\mu    
 \Theta(p^\mu d\hat{\sigma}_\mu) \cos \theta_{\vec{p}}
   f_{int}(x,\vec{p})
\eeq{kin-4}
and
\beq
d T_i^{\mu\nu} = 
      -\frac{dx}{\lambda}\int\frac{d^3p}{p_0} p^\mu p^\nu 
 \Theta(p^\mu d\hat{\sigma}_\mu) 
\cos \theta_{\vec{p}}
f_{int}(x,\vec{p}).
\eeq{kin-5}
If we do not have collision or relaxation terms in our transport equation
then the conservation laws are trivially satisfied. If, however, collision or
relaxation terms are present, these contribute to the change of $T^{\mu\nu}$
and $N^\mu$, and this should be considered in the modified 
distribution function $f_{int}(x,\vec{p})$.

\subsection{Immediate re-thermalization limit}

We can get from the equation (\ref{kin-2}) the general solution  for
$f_{int}(x,\vec{p})$ in the RFF frame:
$$
 f_{int}(x,\vec{p})=e^{-\frac{x}{\lambda}[\Theta(p^{x})\frac{p^{x}}{p}
 +\frac{\lambda}{\lambda'}]}\left\{f_{int}(0,\vec{p}) + 
\phantom{\int_0^x e^{\frac{x}{x}}}  
\right. 
\nonumber
$$
\beq
\left.
 +\frac{1}{\lambda'}\int_{0}^{x}f_{eq}(x',\vec{p})
e^{\frac{x'}{\lambda}[\Theta(p^{x})\frac{p^{x}}{p}
+\frac{\lambda}{\lambda'}]} dx'\right\}\,. 
\eeq{kin-6}

As a first approximation to this solution let us assume 
that $\lambda' \ll \lambda$, i.e. re-thermalization is much faster
than particles freezing out, or much faster than parameters,
$n_{eq}(x)$, $T_{eq}(x)$ and $u^\mu_{eq}(x)$ change.
Eq. (\ref{kin-6}) then leads to 
\beq
f_{int}(x,\vec{p})\approx f_{eq}(x,\vec{p}), \ \ {\rm for} \ \ 
\lambda'\ll \lambda\,.
\eeq{kin-7}

For $f_{eq}(x,\vec{p})$ we can assume the spherical J\"uttner form at
any $x$ including both positive and negative momentum parts 
with parameters $\hat{n}(x),\ T(x)$ and $u_{RFG}^\mu(x)$, where 
$u_{RFG}^\mu(x)$ is the peak velocity of the J\"uttner gas (which 
is the same as the flow velocity of the non-cut J\"uttner gas), and
$\hat{n}(x)$ is the proper density i.e. the density in the frame moving with 
$u_{RFG}^\mu(x)$ \cite{AC98,ALC98}.


\begin{figure}[htb]
\vspace*{-1.0cm}
	\insertplot{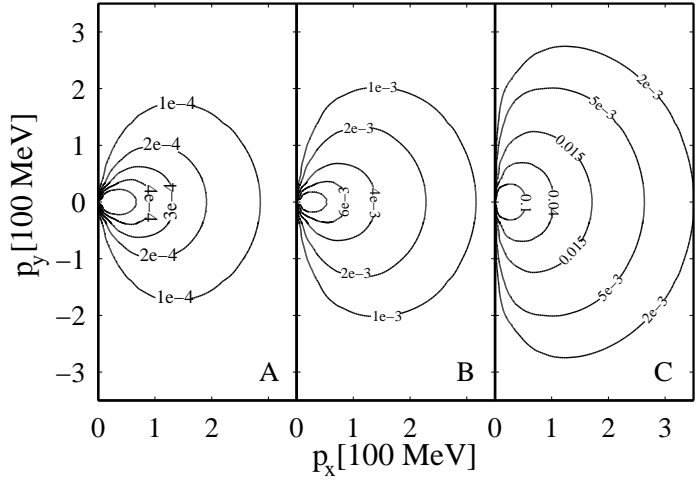}
\caption[]{The freeze-out distribution,
$f_{free}(x,\vec{p})$, in the rest frame of the
freeze out front (RFF) calculated 
according to eqs. (\ref{simmod}). A, B and C correspond
to  $x = 0.2 \lambda,\ 3 \lambda,\  100 \lambda$  respectively and
$u_{RFG}^\mu|_{x=0}=(1,0,0,0)$.
The numbers in the contours are in arbitrary units. The distribution is
asymmetric and elongated in the freeze out 
direction, $x$. This may lead to a large-$p_t$ enhancement, compared 
to the usual J\"uttner assumption used in most previous
calculations as a freeze out distribution.
Note that $f_{int}(x,\vec{p})$ 
does not tend to the cut J\"uttner 
distribution in  the limit $x \rightarrow \infty$.
}

\label{fig:3}
\end{figure}

In this case the change of conserved quantities due to particle drain or 
transfer can be  evaluated for an infinitesimal $dx$. 
We assume that the 3-flow is normal to the
freeze out surface. The changes of the conserved particle currents and
energy-momentum tensor in the RFF, using the notation of ref. \cite{AC98},$^b$
eqs. (\ref{kin-4},\ref{kin-5}) are given by 

\beq
\begin{array}{rll} 
dN_i^0 =
&-\frac{dx}{\lambda}\frac{\tilde{n}}{4v^2\g^2}
\left\lbrace
aK_1(b)+jb(3v^2-1)\g^2[(1+j)K_1(a)-{\cal K}_1(a,b)]
\right.
\nonumber\\ & & \\
&\left.+j\g v^2b^2[(1+j)K_0(a)-{\cal K}_0(a,b)]+2jv^3\g^3(b+1)e^{-b}
\right\rbrace
\nonumber\\ & & \\ 
\stackrel{\scriptscriptstyle m=0}{\longrightarrow}
& -\frac{dx}{\lambda}\frac{\tilde{n}}{4}\g(1+u)^{2},
\nonumber\\ & & \\ 
dN_i^x
=&-\frac{dx}{\lambda}\frac{\tilde{n}}{4jv^3\g^3}
\left\lbrace
-2K_0(b)+2jv\g^2e^{-b}[v^2\g^2(b+1)-v^2b-1]\right.
\nonumber\\ & & \\  
&\left.+j(2+v^4\g^2b^2)[(1+j)K_0(a)-{\cal K}_0(a,b)]\right.+
\nonumber\\ & & \\ 
&\left.+jv^2(3v^2-1)\g^3b[(1+j)K_1(a)-{\cal K}_1(a,b)]
\right\rbrace
\nonumber\\ & & \\
\stackrel{\scriptscriptstyle m=0}{\longrightarrow}
& -\frac{dx}{\lambda}\frac{\tilde{n}}{4u^3\g^3}\left[-2u\g^2(1+u)
-2ln(1-u)+u^{2}\g^{4}(1+u)^{2}\right],
\end{array}
\eeq{dni}
and by the expressions
\beq
\begin{array}{rll}
dT_i^{00}=
&-\frac{dx}{\lambda}\frac{\tilde{n}T}{4v^2\g^2}
\left\lbrace\frac{a}{\g}K_1(b)+jv\g^2e^{-b}\!
\left[ (1{+}3v^2)\g^2\!A(b)-
\right.\right.\nonumber
\\ & & \nonumber\\
&\left.-(2+v^2b^2)(b+1)+v^4(1+\frac{v^2}{3})\g^2b^3\right]+a^2K_0(b)\nonumber
\\ & & \nonumber\\
&+jv^2\g^2b^2(3+v^2)\left[(1+j)K_2(a)-{\cal K}_2(a,b)\right]\nonumber
\\ & & \nonumber\\
&+j(v^2b^2-v^2-1)\g b\left[(1+j)K_1(a)-{\cal K}_1(a,b)\right]\nonumber
\\ & & \nonumber\\
&\left.-jb^2\left[(1+j)K_0(a)-{\cal K}_0(a,b)\right]
\right\rbrace\nonumber
\\ & & \nonumber \\
\stackrel{\scriptscriptstyle m=0}{\longrightarrow}
&  -\frac{dx}{\lambda}\frac{\tilde{n}T}{4u^2\g^2}
\left[1-u^2+2u\g^{4}(1+u)^{3}-\g^{2}(1+u)^2\right]
\ ,\nonumber
\\ & & \nonumber\\
dT_i^{0x}=
&-\frac{dx}{\lambda}\frac{\tilde{n}T}{4jv^3\g^3}
\left\lbrace j(1+3v^2)v^{2}\g^{3}b^2
\left[(1+j)K_2(a)-{\cal K}_2(a,b)\right]\right.\nonumber
\\ & & \nonumber\\
&+jv^2\g^5e^{-b}\left[v(v^2+3)A(b)-a^{2}bv^3+(1+3v^2)\frac{(bv)^3}{3}
-va^2\right]\nonumber
\\ & & \nonumber \\
&\left.+jv^{4}\g^{2}b^{3}\left[(1+j)K_1(a)-{\cal K}_1(a,b)\right]
\right\rbrace - \frac{2T}{jv\g}dN_i^0 \nonumber
\\ & & \nonumber \\
\stackrel{\scriptscriptstyle m=0}{\longrightarrow}
& -\frac{dx}{\lambda}\frac{\tilde{n}T}{2}\g^{2}(1+u)^3 
 \ ,
\\ & & \\
dT_i^{xx}=& -\frac{dx}{\lambda}\frac{\tilde{n}T}{4v^{4}\g^4}
\left\lbrace 
jv^{4}(3+v^2)\g^{4}b^2\left[(1+j)K_2(a)-{\cal K}_2(a,b)\right]\right.
\\ & & \\
&+jv^6\g^3b^3\left[(j+1)K_1(a)-{\cal K}_1(a,b)\right]+
jv^3\g^6e^{-b}
\\ & & \\
&\left.  	
\left[
\frac{v(3+v^2)(bv)^3}{3}+
a^2(v^4b-1)+(3v^2+1)A(b)
\right]	
\right\rbrace
-\frac{3T}{jv\g}dN_i^x 
\nonumber\\ & & \\ 
\stackrel{\scriptscriptstyle m=0}{\longrightarrow}
&-\frac{dx}{\lambda}\frac{\tilde{n}T}{4u^4\g^4}\left[
 2u^3\g^6(1+u)^3-3u^2\g^4(1+u)^2+\right.
\\ & & \\
&\left.+6u\g^2(1+u)+6ln(1-u)\right] \nonumber\ ,
\\ & & \\
dT_i^{yy}=& -\frac{dx}{\lambda}\frac{\tilde{n}T}{8v^4\g^4}
\left\lbrace 
-jv^2\g^3(v^2+1)b\left[(1+j)K_1(a)-{\cal K}_1(a,b)\right]\right.
\\ & & \\
&-jv^4\g^2b^2\left[(1+j)K_0(a){-}{\cal K}_0(a,b)\right]
+bv^2K_1(b)
\\ & & \\
&\left.-2jv^3\g^4(b{+}1)e^{-b}
\right\rbrace+\frac{3T}{2jv\g}dN_i^x 

\\ & & \\
\stackrel{\scriptscriptstyle m=0}{\longrightarrow}
& -\frac{dx}{\lambda}\frac{\tilde{n}T}{8u^4\g^4}
\left[
2u^2\g^4(1{+}u)^2-6u\g^2(1{+}u)+u^2-2ln(1{-}u) \right],
\end{array}
\eeq{dti}

and $dT_i^{zz} = dT_i^{yy}$.
Note that in RFF the flow velocity of the re-thermalized component is
$u_{i,RFG}^\mu(x) = \gamma(x)\ (1,u(x),0,0)|_{RFF}$, where
$\gamma= 1/\sqrt{1-u^2}$; and we also use the notation $v=|u|$ and $j=sgn(u)$.

The new parameters of distribution $f_{int}$, 
after moving to the right by $dx$ can be obtained from
$dN_i^\mu$ and $dT_i^{\mu\nu}$.
The conserved particle density of the re-thermalized spherical
J\"uttner distribution after a step $dx$ is: 
$$
\hat{n}_i(x+dx) = \hat{n}_i(x) + d\hat{n}_i(x)
 = 
\sqrt{ N_i^\mu(x+dx) N_{i,\mu}(x+dx) } \ ,
$$
where the expressions are invariant scalars.
The differential equation
describing the change of the proper particle density is:
\beq
d\hat{n}_i(x) =  u_{i,RFG}^\mu(x)\ dN_{i,\mu}(x) \,. 
\eeq{dnx}
Although this covariant
equation is valid in any frame, we can
calculate it in the RFF, where 
the values of $dN_i^\mu$ were given above in eq. (\ref{dni}).

For the re-thermalized interacting component
Eckart's flow velocity is the velocity of the RFG, which changes with $x$,
so we denote this frame as RFG$(x)$.
The velocity of this frame decreases with decreasing $x$ due to the
particle drain at positive momenta.
For the spherical J\"uttner distribution  
the Landau and Eckart flow velocities are the same,
$
u^\mu_{i,E,RFG}(x) = u^\mu_{i,L,RFG}(x) 
= u^\mu_{i,RFG}(x) 
$.
Thus we can evaluate the flow velocity $u_{i,RFG}^\mu(x+dx)$:
$$
u^\mu_{i,RFG}(x+dx)  =    N_i^\mu (x+dx) / \sqrt{ N_i^\mu N_{i,\mu} }\,,
$$
which leads to the following covariant  expression
\beq
du^\mu_{i,E,RFG}(x)=\Delta^{\mu\nu}_i(x)\  \frac{dN_{i,\nu}(x)}{\hat{n}_i(x)}\,,
\eeq{duex}
where
$
\Delta^{\mu\nu}_i(x) = g^{\mu\nu} -  u_{i,RFG}^\mu(x)\, u_{i,RFG}^\nu(x)\, 
$\ \ 
is a projector to the plane orthogonal to
$
u^\mu_{i,RFG}(x)  
$.
This equation is valid in 
any reference frame, nevertheless we know the four-vectors
on the r.h.s. in the RFF explicitly.
Then the new Eckart flow velocity of the matter 
is 
$
u^\mu_{i,E,RFG}(x+dx) $ $ = $ $  
u^\mu_{i,RFG}(x)$ $ + $ $  du_{i,E,RFG}^\mu(x) 
$.

\begin{figure}[htb]
\vspace*{-1.0cm}
	\insertplot{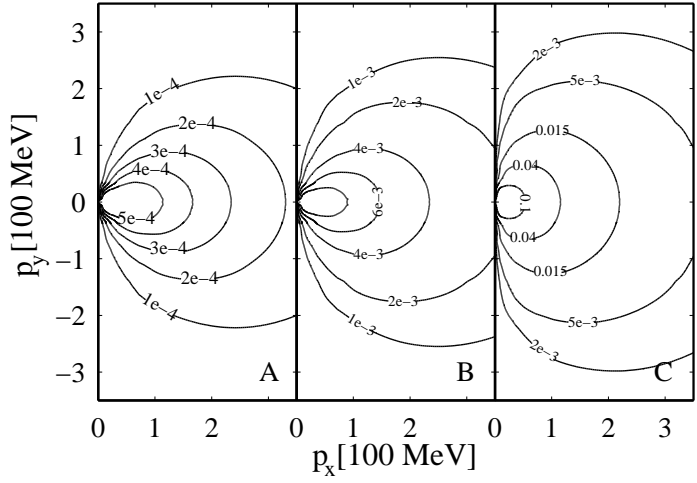}
\caption[]{The same as Fig.~\ref{fig:3}, just 
$u_{RFG}^\mu|_{x=0}=\gamma(1,0.5,0,0)$. We can see that particles are freezing
out faster if we start with positive initial flow velocity.
}

\label{fig:4}
\end{figure}

To get the temperature and the change of Landau's flow velocity, we 
analyze the change of the energy momentum tensor.
Before the particle drain the energy - momentum tensor at $x$ in the RFG is 
diagonal, 
$T^{\mu\nu}_i(x) = {\rm diag}(e_i,P_i,P_i,P_i)|_{RFG(x)}$
while in the RFF 
$T^{\mu\nu}_i(x) $ $=$ $ 
\left[ (e_i + P_i) \right.$ $ u_{i,RFG}^\mu u_{i,RFG}^\nu(x) 
$ 
$ 
\left. - P_i g^{\mu \nu}\right] |_{RFF(x)}
$.
Adding to this the
drain terms, $dT^{\mu\nu}_i(x)$, arising from the freeze out while
we move to the right  by $dx$, yields 
$T^{\mu\nu}_i(x+dx)$ 
which will not be diagonal in the RFG$(x)$ and the 
pressure part will not be isotropic. We can Lorentz transform
this to another frame which diagonalizes 
$T^{\mu\nu}_i(x+dx)$.
This means to find the Landau flow velocity of the new system,
$u^\mu_{i,L,RFG}(x+dx)$ in the original RFG$(x)$. After
a straightforward diagonalization, a somewhat tricky algebra and 
neglecting second and higher order terms
we arrive at the covariant expression\cite{ALC98}
\beq
du^\mu_{i,L,RFG}(x) = \frac{ 
\Delta^{\mu\nu}_i(x)\ \ dT_{i,\nu\sigma}\ \ u^\sigma_{i,RFG}(x)}{e_i + P_i}.
\eeq{dulx}
Although, 
for the spherical J\"uttner distribution  
the Landau and Eckart flow velocities are the same,
the change of this flow velocity  calculated from the baryon current
and from the energy current are different in general
$$
du^\mu_{i,E,RFG}(x) \ne  du^\mu_{i,L,RFG}(x) \,.
$$
This is a clear consequence of the asymmetry caused by the freeze out
process as it was discussed in ref. \cite{ALC98}.


\begin{figure}[htb]
	\insertplot{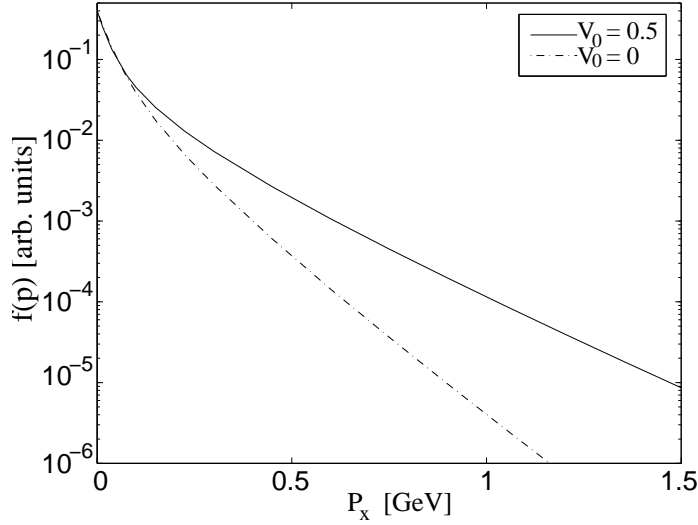}
\vspace*{-0.6cm}
\caption[]{The 
local transverse momentum (here $p_x$)
distribution for baryon free, massless gas 
at $p_y = 0$, $x=100 \lambda$ and $T_0 = 130\,$MeV.  
The transverse momentum spectrum is obviously curved
due to the freeze out process, particularly for large
initial flow velocities. The apparent slope parameter
increases with increasing transverse momentum. This behavior agrees
with observed pion transverse mass spectra at SPS \cite{na44,na49}.
}

\label{fig:5}
\end{figure}

In the special case of the {\em massless limit} we can calculate 
$du^\mu_{i,E,RFG}(x)$ and  $du^\mu_{i,L,RFG}(x)$ in the RFG. 
In this frame $u^\mu_i=(1,0,0,0)$, and according to (\ref{duex})
and (\ref{dulx}) we can get:
 \mbox{$du^0_{i,E} = du^0_{i,L}=0$, $du^y_{i,E} = du^y_{i,L}=0$,
  $du^z_{i,E} = du^z_{i,L}=0$} and
$$
du^x_{i,E} = \frac{dN^x_{i}}{n}= \frac{dN^x_{i}}{N^0_{i}}
$$
$$ 
du^x_{i,L} =  \frac{dT^{x0}_i}{e+P}= \frac{dT^{x0}_i}{T^{00}_i+T^{xx}_i}.
$$
Here the last values of both equations are in the RFG.
Thus we can see that
\mbox{$du^x_{i,E} = {4\over 3} du^x_{i,L}  \,,$}
in the massless limit.
Using the definitions of the $dN^x$ and $dT^{x0}$ from equations 
(\ref{kin-4}),(\ref{kin-5}) and $N^0$ and $T^{00}$ 
from section \ref{three} \ \ , it is easy to 
see, that in the limit $m \rightarrow 0$, (and $p^0=p$) the calculation
leads to the integral:
$$
du^x_{i,E} = 
    {4\over 3} du^x_{i,L} = 
   -\frac{dx}{2\lambda}\int_{-1}^{1}d\cos\theta\ \
\frac{\cos\theta (\cos\theta-u)}{1-u\cos\theta}
\Theta(u-\cos\theta)
\,,
$$
where $u$ is defined via $u_{i,RFG}^\mu = \gamma(x) (1, u, 0, 0)$ (in RFF),
and $\cos \theta$ is the polar angle of the emitted particle in the RFG.
(The polar emission angle in RFF, given earlier, can be expressed as 
$
\cos \theta_{\vec{p}} = (\cos\theta - u)/(1-u \cos\theta)\,.
$)

This problem does not occur for the freeze out of baryonfree 
plasma, and we have only $du^\mu_{i,L}$. 

The last item is to determine the change of the temperature parameter
of $f_{int}$.  From the relation
$
e\equiv u_\mu T^{\mu\nu} u_\nu
$
we readily obtain the expression for the change of energy density
\beq 
de_i(x) =u_{\mu,i,RFG}(x)\ dT^{\mu\nu}_i(x)\ u_{\nu,i,RFG}(x) \,,
\eeq{dedx}
and from the relation between the energy density and the
temperature (e.g. Chapter 3 in ref.~\cite{Cs94}), we can obtain
the new temperature at $x+dx$.
Fixing these parameters we fully determined the spherical J\"uttner
approximation for $f_{int}$.

The application of this model to the baryonfree and massless gas gives
 the following coupled set of equations:
\beqar
& \frac {d\ln{T}}{dx}=-\frac{u_{\mu}\tau^{\mu\nu}u_{\nu}}{4\sigma_{SB}}\,, 
\nonumber & \\ & & \\
& \frac {du^{\mu}}{dx}=-\frac{3}{4\sigma_{SB}}\left[\tau^{\mu\nu}-u^{\mu}
 u_{\sigma}\tau^{\sigma\nu}\right]u_{\nu}\,. \nonumber &
 \eeqar{simmod}
Here we use the EoS, $e=\sigma_{SB} T^4$,\ \ and the definition 
\mbox{$dT^{\mu\nu}$ $=$ $-dx\,\tau^{\mu\nu} T^4$,}\ \  
where $dT^{\mu\nu}$ are given by eqs.
(\ref{dti}), and so $x$ is measured in units of $\lambda$.

The results of numerical calculation are displayed in 
Figs.~\ref{fig:2} - \ref{fig:5}.  
The velocity and temperature of the interacting component are gradually
decreasing due to the loss of particles carrying most momentum and thermal
energy. Thus, we see that $T \rightarrow 0$, when $x \rightarrow
\infty$. So, $f_{int}(x,\vec{p})=\frac{1}{(2\pi\hbar)^3}
\exp[(\mu-p^{\nu}u_{\nu})/T] \rightarrow 0$, when $x \rightarrow \infty$. 
Thus, all particles freeze out in the model with rescattering.

Now we can find the distribution function for the noninteracting, frozen out
part of particles according to equation (\ref{kin-3}). The results are shown
in Fig. \ref{fig:3} - Fig. \ref{fig:5}.  We would like to
note that now $f_{free}(x,\vec{p})$ does not tend to the cut J\"uttner
distribution in the limit $x \rightarrow \infty$, it has a smooth,
anisotropic shape. Most importantly the slope in the FO direction is
not exponential, resembling recent experimental data.

\section{Volume emission model}
\label{subfour}

In this section we demonstrate an improvement which yields complete FO in a
layer of finite width.  We calculate the kinetic freeze out distribution
based on four volume emission models \cite{GH95,GH96,GH97}. In order to
illustrate the physical mechanism of this freeze out process, let us study
again a simple one-dimensional flow \cite{AC98,ALC98}. We suppose an infinite
tube where a stationary flow of a fluid is supplied from the left ($x<0$), so
that the freeze out occurs for the positive direction of $x$. 

In the four volume emission model, we introduce the basic quantity, the
so-called escape probability
\begin{equation}
{\cal P}(\vec{r},t,\vec{p})\equiv e^{-  \int_t^\infty \sigma v_{rel}
   {\sf n}(\vec{r}+\vec{v}t,t)dt},  
 \label{P}
\end{equation}
where ${\sf n}$ is the total density in the calculational frame, 
$\vec{v}=\vec{p}/E$, the velocity of the
particle,
$\sigma $,  the  total cross section and $v_{rel}$, the relative velocity.
 
We can understand this assumption in the following way: let us think
for simplicity  of an
ideal gas of hard balls with radius $R_{eff}$, so that the collision cross
section becomes $\sigma_0 = 4\pi R_{eff}^2$. A particle is frozen out at time
$t$ if it will not collide any more starting from this time.  The probability
to have no collisions, or to freeze out, is described by a Poisson
distribution
\begin{equation}
P = W(0) = e^{-\rho N} \,,
\end{equation}
where $\rho$ is the probability of a collision, and $N$
is the total number of particles. Then, in the hard
ball approximation  $\rho N$ is the number of particles 
which our particle can meet on its way
 \begin{equation}
\rho N = \int_t^\infty \sigma_0 v_{rel} {\sf n}(\vec{r}+\vec{p}/Et,t) dt \,,
\end{equation}
where ${\sf n}(x)$ is the particle density in the calculational frame
 ${\sf n}(x)=N^0(x)$.
This leads us to  the expression (\ref{P}), where the interaction 
can be included through hard ball cross section $\sigma=\sigma_0$ or 
the effective cross section $\sigma=\sigma_{eff}$. 

For a stationary one dimensional case, we can express the escape probability,
$\cal{P}$, as$^c$

\beq
{\cal P}(\vec{r},t,\vec{p}) \rightarrow {\cal P}(x,\cos \theta) = 
\left\{ \begin{array}{ll}
e^{-   \int_x^\infty \sigma    {\sf n}(x)\frac{dx}{\cos \theta}} \,, \quad 
\cos \theta \geq 0 \nonumber \\ \ \ \\ 
e^{-   \int_x^{-\infty} \sigma    {\sf n}(x)\frac{dx}{\cos \theta}} \,, \ \
\cos \theta \leq 0 \nonumber \end{array} \right.
\eeq{Px1}
where 
$
\cos \theta =\frac{p^x}p 
$.
Let us rewrite this result in the following form
\beqar
{\cal P}(x,\cos \theta) = e^{-\sigma N \frac{\Theta(\cos \theta)}{\cos \theta}
+ \frac{\sigma}{\cos \theta} \int_{-\infty}^x {\sf n}(x)dx} \,,
\eeqar{Px}
where $N = \int_{-\infty}^{+\infty} {\sf n}(x) dx$ is the total number of particles.
Note that $N$ is so large that 
$exp\left(-\sigma N\right) \rightarrow 0$.  
Let us check different cases. We have
\begin{eqnarray}
\cos \theta \geq 0 \quad \Rightarrow \quad {\cal P}(-\infty,\cos \theta) 
\rightarrow 0 \,,
 \quad {\cal P}(\infty,\cos \theta) = 1 \,,
 \nonumber \\ & & \\ 
 \cos \theta \leq 0 \quad \Rightarrow \quad {\cal P}(-\infty,\cos \theta) = 1 
 \,,
 \quad {\cal P}(\infty,\cos \theta) \rightarrow 0 \,,
\nonumber 
\end{eqnarray} 
The result for $cos \theta \leq 0$, $x=-\infty$ seems to be a little confusing, 
but it just shows that a particle going in negative direction and starting from 
$x=-\infty$ has no particle to collide with. We are interested in the 
region with positive $x$: 
 \begin{eqnarray}
 {\cal P}(0,\cos \theta) = e^{- N \sigma \frac{\Theta(\cos \theta )}{\cos 
 \theta}
  + N' \frac{\sigma}{\cos \theta}} \,,
\end{eqnarray} 
where $N' = \int_{-\infty}^0 {\sf n}(x) dx$. Let us assume that $N$ and $N'$
are such that \linebreak $exp\left(-\sigma N'\right) \rightarrow 0$ and
$exp\left(-\sigma (N-N')\right) \rightarrow 0$. In this case 
\beq 
{\cal P} (0,\cos \theta) \rightarrow 0,\quad \forall \cos \theta \,.  
\eeq{to0} 
This condition is easily satisfied , because there is always a point $A$ such
that $\int_{-\infty}^A {\sf n}(x) dx = \frac{N}{2}$, and then we should shift
origo of our frame to this point along the $x$ axis.

This escape probability determines the free particle distribution 
$f_{free}(x,p)$ as a
fraction of the total particle distribution
\beq
f_{free}(x,p)={\cal P}f(x,p)\,,
\eeq{eqfree}
and the interacting part of the particle distribution $f_{int}$ is defined as
\beq
f_{int}(x,p)=(1-{\cal P})f(x,p)\,,
\eeq{eqint}
where 
$
f(x,p)=f_{free}(x,p)+f_{int}(x,p)\,.
$
 The total
density ${\sf n}(x)$ is given as 
\begin{equation}
N^0(x)={\sf n}(x)=\int d^3pf(x,p).
\end{equation} 
Note that for
$x \rightarrow -\infty$, ${\sf n}(x) \rightarrow constant$. Condition
(\ref{to0}) means that we do not have frozen out particles at $x=0$, i.e.,
$f(0,p)=f_{int}(0,p)$.  It is obvious from eq.(\ref{Px}) that if ${\sf n}$ is
constant, then ${\cal P}$ becomes identically zero, because $N \rightarrow
\infty$, and the post-FO component can never emerge. Another important fact that
could be seen from eq.(\ref{Px}) is that if  
 $$
{\cal P}(0,\cos \theta) = 0 \,, \quad \cos \theta \leq 0
 $$
it will be equal $0$ for all $x > 0$. So backward going particles can not
freeze out in our consideration. From this point of view this volume emission
model close to idealized model with drain term discussed in section
\ref{six}. \ We will see later that the results of these calculations are
similar in some aspects to those in the kinetic freeze out 
models discussed above.  

If the system is truly one-dimensional for all $x$ values, then the total
density $n $ should vanish for large $x$, otherwise ${\cal P}$ vanishes.
However, this contradicts the assumed conservation of flux in the stationary
case:
\begin{eqnarray}
N^1(\infty)=N^1(0)=\int d^3p (p^x/E) f(\infty,p) \neq 0 \nonumber
\end{eqnarray}
is incompatible with 
\begin{eqnarray}
N^0(\infty) ={\sf n}(\infty ) = \int d^3p  f(\infty,p) = 0 \nonumber \,,
\end{eqnarray}
since the velocity $ p^x/E \leq 1$.  Therefore, to get a stationary
one-dimensional flow the system should have a finite size in the freeze out
direction.

Suppose that there exists a boundary at $x=L>0$, so that for $x>L$, the
density falls off very rapidly and the escape probability is almost zero
there.  Such a situation happens for a semi-infinite tube open to the vacuum
at $x=L$.  We should write 
\begin{equation}
N=  \int_{-\infty}^L {\sf n}(x) dx \,.
\end{equation}

We are going to show that the equations (\ref{eqfree}, \ref{eqint})
together with the
conservation laws determine all the distribution functions when we assume a
thermal spectrum for the interacting component.

First, note that all  distributions are specified if,

\begin{enumerate}
\item  the interacting flow velocity, $v_{int}(x)$,

\item  the interacting temperature, $T(x)$,

\item  the interacting density, ${\sf n}_{int}(x),$ and

\item  the escape probability in the $x$ direction, 
\begin{equation}
P_0(x)\equiv e^{\int_{0}^x  \sigma {\sf n}(x)dx},  
\label{P0}
\end{equation}
\end{enumerate}

\noindent are known. To see this, first we write 
\begin{equation}
{\cal P}(x,\cos \theta )= P_1(\cos \theta)\left\{ P_0(x)\right\}^{\frac{1}
{\cos \theta }},
\label{PP0}
\end{equation}
where
\beq
P_1(cos \theta)={\cal P}(0,\cos \theta )=e^{-\sigma N \frac{\Theta(\cos \theta)}{\cos \theta}+ 
N' \frac{\sigma}{\cos \theta}} \,,
\eeq{defP1}
and express the total and free distributions in terms of $f_{int}$. 
\begin{eqnarray}
f(x,p) &=&\frac 1{1-{\cal P}}f_{int}(x,p)=\frac 1{1-{\cal P}}{\sf n}_{int}(x)
\frac 1Z e^{-p^\mu u^{int}_\mu /T},  \label{ftot} \\
f_{free}(x,p) &=&\frac{{\cal P}}{1-{\cal P}}{\sf n}_{int}(x)\frac 1Z
e^{-p^\mu u^{int}_\mu /T},  \label{fFO}
\end{eqnarray}
where 
\[
u^{int}_\mu =\left( 
\begin{array}{c}
\gamma \\ 
-\gamma v_{int}
\end{array}
\right) , 
\]
with $\gamma =1/\sqrt{1-v_{int}(x)^2}$ and $Z$ is the normalization factor, 
\[
Z=Z(T)=\int d^3p\ e^{-p^\mu u^{int}_\mu /T}. 
\]
In our stationary regime, the conservation laws are expressed as 
\begin{equation}
N^1(x) = const = N^1(0), \ \ 
T^{01}(x) = const = T^{01}(0), \ \ 
T^{11}(x) = const = T^{11}(0), 
\end{equation}
where 
\[
N^1(x)\equiv \int d^3p\frac{p^x}{p^0}f(x,p), 
\]
\[
T^{01}(x)\equiv \int d^3p\;p^xf(x,p), 
\]
\[
T^{11}(x)\equiv \int d^3p\;\frac{(p^x)^2}{p^0}f(x,p). 
\]
Once the initial values $N^1(0), T^{01}(0)$ and $T^{11}(0)$
are specified, these equations,
together with Eq. (\ref{ftot}), determine algebraically 
$T(x)$, $v_{int}(x)$ and ${\sf n}_{int}(x)$, at each $x$ as functions of $P_0(x)$.

On the other hand, from Eq.(\ref{P0}) 
\begin{equation}
\frac 1{P_0}\frac{dP_0}{dx}= \sigma  {\sf n}(x),  \label{dp0dx}
\end{equation}
and 
\begin{equation}
N^0(x)={\sf n}(x)=\int d^3pf(x,p)={\sf n}_{int}(x)\int d^3p\frac 1{1-{\cal P}}\frac 1Ze^{-p^\mu
u^{int}_\mu /T},
\end{equation}
so that we get an integro-differential equation for $P_0$, 
\begin{equation}
\frac 1{P_0}\frac{dP_0}{dx}=
\sigma   {\sf n}_{int}(x)\int d^3p\frac 1{1-{P_1(\cos \theta )P_0{}^{1/\cos
    \theta }}}\frac
1Ze^{-p^\mu u^{int}_\mu /T}.  \label{diffP0}
\end{equation}

To compute $n=N^0$, $N^1$, $T^{01}$ and $T^{11}$, we need to know the integrals: 
\[
I_1[P_{int},T,v_{int}]\equiv \int d^3p\frac 1{1-P_1(\cos \theta )P_0{}^{1/\cos
    \theta }} e^{-p^\mu u^{int}_\mu /T}, 
\]
\[
I_2[P_{int},T,v_{int}]\equiv \int d^3p\frac{p\cos \theta }{p_0}\frac
1{1-P_1(\cos \theta )P_0{}^{1/\cos \theta }}e^{-p^\mu u^{int}_\mu /T}, 
\]
\[
I_3[P_{int},T,v_{int}]\equiv \int d^3p\frac{p\cos \theta }{1-P_1(\cos \theta )P_0{}^{1/\cos \theta }%
}e^{-p^\mu u^{int}_\mu /T}, 
\]
\[
I_4[P_{int},T,v_{int}]\equiv \int d^3p\frac{p^2\cos ^2\theta }{p_0}\frac
1{1-P_1(\cos \theta )P_0{}^{1/\cos \theta }}e^{-p^\mu u^{int}_\mu /T}. 
\]
Note that these functions are not scalar, and the above expressions
are valid in the frame where the density distributions are at rest, i.e.
in the LR frame.
The local rest frame quantities are labeled by $*$. Then in the local rest 
frame 
$p\rightarrow p^{*}$,
$p^\mu u_\mu \rightarrow p^{*0}=E^{*}$, and:
\[
I_1[P_{int},T,v_{int}]\equiv 
\int d^3p^{*}\frac 1{1-P_1(\cos \theta )\left\{ P_0(x)\right\}
^{1/\cos \theta }}e^{-E^{*}/T},
\] 
\[
I_2[P_{int},T,v_{int}]\equiv \int d^3p^{*}\frac{p^{*}\cos \theta ^{*}}{E^{*}}\frac
1{1-P_1(\cos \theta )\left\{ P_0(x)\right\} ^{1/\cos \theta }}e^{-E^{*}/T}, 
\]
\[
I_3[P_{int},T,v_{int}]\equiv \int d^3p^{*}\frac{p^{*}\cos \theta ^{*}}
{1-P_1(\cos \theta )\left\{
P_0(x)\right\} ^{1/\cos \theta }}e^{-E^{*}/T}, 
\]
\[
I_4[P_{int},T,v_{int}]\equiv \int d^3p^{*}\frac{p^{*2}\cos ^2\theta ^{*}}{E^{*}}%
\frac 1{1-P_1(\cos \theta )\left\{ P_0(x)\right\} ^{1/\cos \theta }}e^{-E^{*}
/T}, 
\]
 with
$$
\cos \theta ={p_x}/p = \frac 
{ \gamma p^{*}\cos \theta ^{*}+\gamma v_{int} E^{*} }
{ \sqrt{E^{*2}-p^{*2}\cos ^2\theta ^{*}+\left( \gamma p^{*}\cos 
\theta ^{*}+\gamma v_{int}E^{*}\right) ^2 } } 
$$
and the limits of integral are restricted by 
$\gamma p^{*}\cos \theta ^{*}+\gamma v_{int} E^{*}\geq 0$.

\subsection{ Extrapolation approach}
\label{extrap} 

It is obvious that solving integro-differential equation (\ref{diffP0}), and
evaluating the above integrals require nontrivial numerical calculations.
Nevertheless, we would like to show a simple approach of the solution of this
model.  We can not get the equations for ${\sf n}_{int}(x) \,, u^\mu (x)$ and
$T(x)$, since we can not evaluate the integrals in the conservation laws,
but we are going to 
extrapolate $f^{free}(x)\,, f^{int}(x)$ and ${\sf n}(x)$. 


\begin{figure}[htb]
\vspace*{-1.0cm}
	\insertplot{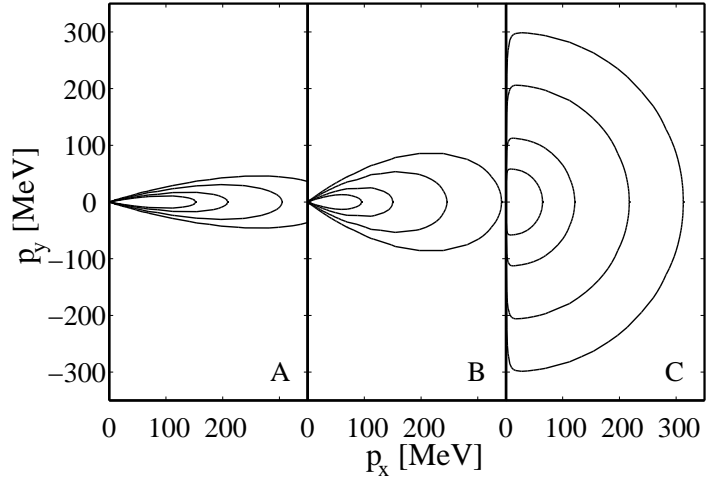}
\caption{\footnotesize 
The freeze out distribution,
$f^{free}(x,\vec{p})$, in the Rest Frame of the
freeze out Front (RFF) calculated 
according to eqs. (\ref{tildeP}-\ref{n1}). A, B and C correspond
to  $x = 30 {\rm fm},\ 80 {\rm fm},\  89.66 {\rm fm} $  respectively and
$u_{RFG}^\mu|_{x=0}=(1,0,0,0)$. We put $N=180 \,, N'=90 \,, m=0\,, n_0=1$ 
and $\sigma={\rm fm}^2=10 \ {\mu}{\rm barn}$.
Contours correspond to $f^{free}=[1,\ 2,\ 4,\ 6] \cdot 10^{-35} (A),
 10^{-13} (B), 10^{-9} (C)\ \frac{fm^{-1}}{\rm MeV^3}$. 
Note that the form of the final freeze out distribution (C)  seems to be 
similar to the cut J\"uttner distribution for large $p_x$. 
}

\label{extf0}
\end{figure}
 
Let us define a set of points on the interval $[0$, $L]$ so that: 
\beq
\{x_i\},i=1,...,m\,, \quad x_i \subset [0,L] \,,\ x_1=0,\ x_m=L \,.
\eeq{setx} 
\beq
\dlt x_i = x_{i+1}-x_i \,, \quad 0 < {\dlt x_i \over L} \ll 1\,.
\eeq{deltax}
We assume that $f_1(p)=f(0,p)= f^{int}_1(p)=f^{int}(0,p)$ is known (i.e. 
${\sf n(0)}\,, u^\mu (0)$ and $T(0)$ are known). The normalization factor 
$Z$ can be evaluated as
\beq
Z(T)\Bigg|_{RFF}=\int d^3p \ e^{-p^\mu u^{int}_\mu /T}= 4 \pi \g T^3 a \left(
2K_1(a) +K_0(a)\right)
\stackrel{\scriptscriptstyle m=0}{\longrightarrow}
8 \pi \g T^3\,. 
\eeq{Zm0}
We have calculated $Z(T)$ in RFF using the same mathematics as in the 
previous section, and all calculations below will be made in RFF too. The
probability $P_i(\cos{\theta})\equiv P(x_i,\cos{\theta})$ is then 
\beq
{\cal P}_i(\cos \theta)= e^{- N \sigma \frac{\Theta(\cos \theta )}{\cos 
 \theta} + N' \frac{\sigma}{\cos \theta}+ 
\frac{\sigma}{\cos \theta}\sum\limits_{j=1}^{i-1} \dlt x_i 
\frac{({\sf n}_i+{\sf n}_{i+1})}{2}} \,,\quad {\cal P}_1(\cos \theta) \equiv 0 \,.
\eeq{acP}
If ${\sf n}(x)$ is a slowly varying  function of $x$, we can extrapolate the 
escape probability by 
\beq
\tilde{\cal P}_{i+1}(\cos \theta) ={\cal P}_i(\cos \theta)
 e^{\frac{\sigma}{\cos \theta} \dlt x_i {\sf n}_i} \,.
\eeq{tildeP}
Since ${\cal P}_i \ll 1$ for almost all values of $x_i$, except for the few
last nodes, $f^{int}$ is a much more smooth function of $x$ than $f^{free}$,
and its extrapolation will be:
\beq
f^{int}_{i+1}(p) = \left( 1 - \tilde{\cal P}_{i+1}(\cos \theta)\right) f_i(p)
 \,.
\eeq{fint1}
Next we calculate:
\beq
\tilde{f}_{i+1}(p) =  f^{int}_{i+1}(p) + f^{free}_{i}(p) \,,
\eeq{tildef}
\beq
\tilde{{\sf n}}_{i+1} =  \int d^3 p \ \tilde{f}_{i+1}(p) \,,
\eeq{tilden}
\beq
{\cal P}_{i+1}(\cos \theta) ={\cal P}_i(\cos \theta)
\ e^{\frac{\sigma}{\cos \theta} \dlt x_i \frac{{\sf n}_i+
\tilde{{\sf n}}_{i+1}}{2}} \,,
\eeq{P1}
\beq
f^{free}_{i+1}(p) = {\cal P}_{i+1}(\cos \theta) \tilde{f}_{i+1}(p) \,,
\eeq{ffree1}
\beq
f_{i+1}(p) =  f^{int}_{i+1}(p) + f^{free}_{i+1}(p) \,,
\eeq{f1}
\beq
{\sf n}_{i+1} =  \int d^3 p f_{i+1}(p) \,.
\eeq{n1}


\begin{figure}[htb]
\vspace*{-1.0cm}
	\insertplot{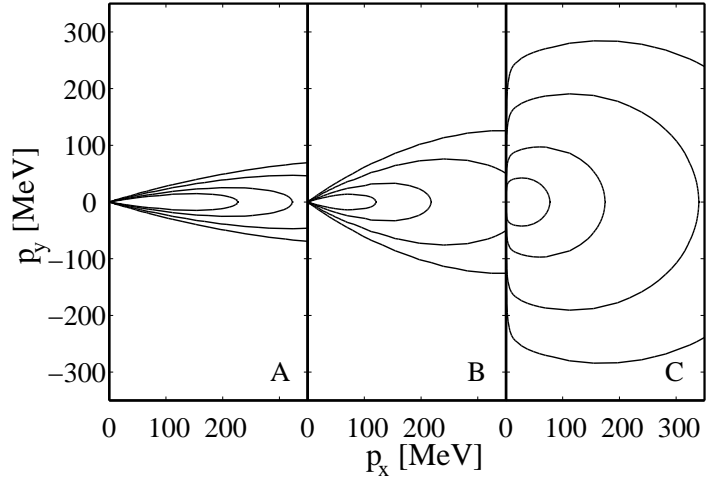}
\caption[]{\footnotesize
The same as Fig.~\ref{extf0}, just for 
$u_{RFG}^\mu|_{x=0}=\gamma(1,0.5,0,0)$.
}

\label{extf05}
\end{figure}

\begin{figure}[htb]

   \insertplot{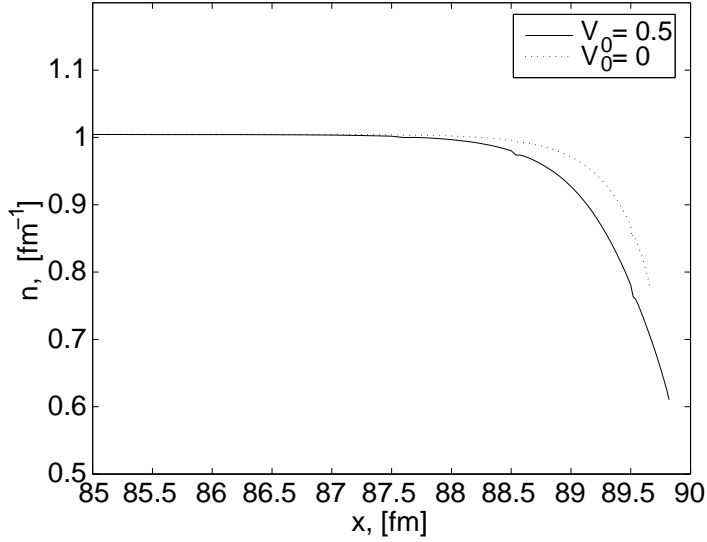}
\caption[]{\footnotesize
Total density ${\sf n}(x)$.We assume $N=180 \,, N'=90\,, m=0\,, n_0=1$ 
and $\sigma={\rm fm}^2=10 \ {\mu}{\rm barn}$.
We can see ${\sf n}(x)$ is almost the same for $x \subset
[0,87] {\rm fm}$, but then it decreases very sharply.  
}

\label{extn}
\end{figure}

Results received according to such an extrapolation are shown in the Figures
\ref{extf0} - \ref{extn}. We can see that ${\sf n}(x)$ 
decreases very sharply and vanishes at point $x=L$, determined from the
condition $N = N' + \int_0^L {\sf n}(x) dx$. We  can also observe the
similarity of Figures \ref{extf0} 
 with the cut
J\"uttner distribution. 
At the same time the initial stages are much more
elongated in the $x$ direction in Figures \ref{extf0} and \ref{extf05}, what
may also lead to a large-$p_t$ enhancement. This model leads to incomplete 
freeze out as the one, presented in section \ref{six}.\ \ \ , but such a problem 
can be cured if rescattering would also be
included in the model, as we demonstrated in the previous case.

So far in the calculations described above we did not take into account 
the conservation laws. Let us check do we break 
conservation laws or not.
According to (\ref{fint1},\ \ref{ffree1},\ \ref{f1}) we have:
\pagebreak
\beqar
f_{i+1}(p)=\left( 1 - \tilde{\cal P}_{i+1}(\cos \theta)\right) f_i(p)+
{\cal P}_{i+1}(\cos \theta)\left[\left( 1 - \tilde{\cal P}_{i+1}(\cos \theta)
\right) f_i(p)+ \right. 
\nonumber \\& & \\
 \left. +{\cal P}_i \left[\left( 1 - \tilde{\cal P}_{i}(\cos \theta)\right) 
f_{i-1}(p)
+{\cal P}_{i-1}\left[\left( 1 - \tilde{\cal P}_{i-1}(\cos \theta)\right) 
f_{i-2}(p)+...\right]\right]\right] \nonumber \,.
\eeqar{recur}
\beq
f_{i+1}(p)-f_i(p) = \left({\cal P}_{i+1}(\cos \theta)-
\tilde{\cal P}_{i+1}(\cos \theta)\right) + O\left({\cal P}^2\right) \,,
\eeq{delf1}
since 
${\cal P}_{i} \ll 1$ 
as well as 
$\tilde{\cal P}_{i} \ll 1$.
Using (\ref{tildeP},\ \ref{P1}) we get:
\beq
f_{i+1}(p)-f_i(p) = 
{\cal P}_i(\cos \theta) 
e^{\frac{\sigma}{\cos \theta}
   \dlt x_i {{\sf n}_i \over 2}}  
\left(
       e^{ \frac{\sigma}{\cos \theta}\dlt x_i 
           { \tilde{{\sf n}}_{i+1} \over 2}      }-
       e^{ \frac{\sigma}{\cos \theta}\dlt x_i  
           { {\sf n}_i \over 2}                }
\right) f_i(p) \,.
\eeq{delf2}


Let us define 
$$
dF = \frac{d^3p}{p^0}p^x \quad {\rm or} \quad \frac{d^3p}{p^0} p^x p^0
\quad {\rm or} \quad \frac{d^3p}{p^0} p^x p^x \,.
$$
Then the conservation laws take the form 
\beq
\int dF\  f_i(p) = Const \,,
\eeq{conlaw}
where $Const = N^1(0) \,,\ {\rm or}\ T^{01}(0)\,,\ {\rm or}\ T^{11}(0)$ 
respectively.
Assume that  this  is true for $f_i(p)$, and then let us check $f_{i+1}(p)$.
\beq
{\cal P}_i(\cos \theta )= P_1(\cos \theta)\ e^{\frac{\sigma {\sf n}_i}{\cos \theta}} 
\,,
\eeq{PNi}
where ${\sf n}_i = \sum \limits_{j=1}^{i-1} \dlt x_i 
\frac{({\sf n}_i+{\sf n}_{i+1})}{2}$, and $P_1(\cos \theta)$ is defined by 
(\ref{defP1}).
The changes of conserved values at the $i$th step of extrapolation
 are given by
\beq 
\int dF \left(f_{i+1}(p){-} f_i(p)\right)  = 
\int   \left(
  1+\frac{\sigma \dlt x_i}{\cos \theta}{{\sf n}_{i} \over 2}
       \right) 
\frac{\sigma \dlt x_i}{\cos \theta} 
{(\tilde{{\sf n}}_{i+1}{-}{\sf n}_i) \over 2}
f_i(p) P_1(\cos \theta)\ e^{\frac{\sigma {\sf n}_i}{\cos \theta}} dF \,. 
\eeq{100}
Since $\tilde{{\sf n}}_{i+1} \approx {\sf n}_{i+1}$ with a good accuracy, and 
${\cal P}_{i}(\cos \theta)$ has its maximum at $\cos \theta = \pm 1$ 
(if we put $N'=
{N \over 2}$ we get ${\cal P}_{i}(1) = {\cal P}_{i}(-1)$) we have 
\beq 
\int dF \left(f_{i+1}(p)- f_i(p)\right) \leq  P_1(1)\ e^{\sigma {\sf n}_i}
{\sigma \over 2} \left(\frac{d{\sf n}}{dx}\right)_i \left(\dlt x_i\right)^2
\cdot Const \,.
\eeq{101}
Finally,
\beq
\int dF \left(f_k(p)- f_1(p)\right) \leq P_1(1)\ Const\  {\sigma \over 2} \ 
\sum \limits_{i=1}^{k-1}e^{\sigma {\sf n}_i}
\left(  
\frac{d{\sf n}}{dx}
\right)_i 
\left(\dlt x_i\right)^2 \,.
\eeq{102}
We have seen that $\left(\frac{d{\sf n}}{dx}\right)_i \ll 1$ (Fig.
\ref{extn}) and
${\cal P}_{i}(1)\ll 1$ for all $i$, except the last few nodes. So, making
$\dlt x_i$ small enough, we can keep $dN^x,\ dT^{0x},\ dT^{xx}$ conserved
with the necessary accuracy.

\section{Conclusions}

In this work we evaluated in a simple kinetic model the freeze out
distribution, $f_{free}(x,p)$, for stationary freeze out across a surface
with spacelike normal vector, $d\hat{\sigma}^\mu d\hat{\sigma}_\mu < 0$.

The first simple kinetic freeze out model, adopted from \cite{ALC98} (see
section \ref{six}.\ ) reproduces the cut J\"uttner distribution as the
limiting distribution, $f_{free}$, after complete freeze out at large
distances. However, the model at the same time leads to unrealistic
consequences, namely that the interacting part of the distribution, $f_{int}$
also survives fully, as the other part of the J\"uttner distribution.  Thus,
having both components at the end in this model, the physical freeze out is
actually not realized.

Here we have presented a solution for an improved but still rather
approximate kinetic freeze out model which takes rescattering into account
(see section \ref{seven}\@.\phantom{5.}\ ).
In this model the interacting component is assumed to be instantly
re-thermalized taking a spherical J\"uttner shape at each time step with
changing parameters.  The three parameters of the interacting component,
$f_{int}$, are obtained in each time step.  The density of the interacting
component gradually decreases and disappears, the flow velocity also
decreases and the energy density decreases also.  The temperature, as a
consequence of the gradual change in the emission mechanism, 
gradually decreases at the
final stages of the freeze out, when only high energy, forward going particles
are taken away from the interacting component.

The arising post freeze out distribution, $f_{free}$ is a 
superposition of cut J\"uttner type of components, from a series of gradually
slowing down J\"uttner distributions. This leads to a 
final momentum distribution with a more dominant peak
and a forward halo, Fig. \ref{fig:5}.  
In this rough model a large fraction ($\sim 95\%$)
of the matter is frozen out by $x=3 \lambda$, thus the distribution
$f_{free}$ at this distance can be considered as a first estimation
of the post freeze out distribution.  One should also keep in mind that 
the model presented here  does not have realistic behavior in the limit 
$x \longrightarrow \infty$, due to its one dimensional character.
Nevertheless, this improved model with rescattering enables
complete freeze out.

The second model presented here (see section \ref{subfour}\@.\phantom{5.}\ )
 shows that we can achieve full freeze out in
finite length even in oversimplified one dimensional models. Thus, the
drawback of the previous model that full freeze out happened exponentially
slowly, can be remedied by including more realistic features in the model in
a straightforward way.

These studies indicate that more attention should be paid to the final freeze
out process, because a more realistic freeze out description may lead to
large $p_t$ enhancement \cite{na44,na49} as the considerations above indicate
(Fig. \ref{fig:5}).  If the heavy ion reaction is basically described by a
kinetic model of weakly interacting hadrons, then idealized FO models with an
assumed FO hypersurface are on the limits of their applicability, and even
then, such models could only be applicable for the heaviest systems. 

If, however, QGP is formed in heavy ion reactions, the number of degrees of
freedom increases tremendously, and continuum models become more suitable to
the problem than dilute gas kinetic or string models with binary
interactions.  In case of rapid hadronization of QGP and simultaneous freeze
out, the idealization of a freeze out hypersurface may be justified, however,
an accurate determination of the post freeze out hadron momentum distribution
would require a nontrivial dynamical calculation.

\section*{Acknowledgement}
This work is supported in part by the Research Council of Norway, PRONEX
(contract no. 41.96.0886.00), FAPESP (contract no. 98/2249-4) and CNPq.
Cs. Anderlik, L.P. Csernai and Zs.I. L\'{a}z\'{a}r are thankful for the
hospitality extended to them by the Institute for Theoretical Physics of the
University of Frankfurt where part of this work was done. L.P. Csernai is
grateful for the Research Prize received from the Alexander von Humboldt
Foundation.
\medskip

\hfill {\it Dedicated to the memory of V.N. Gribov.}

\section*{Notes {\small}} 
\begin{notes}
\item[a]
For the classification scheme see http://www.aip.org/pubservs/pacs.html.

\item[b]
Assuming that the matter is characterized with 4-velocity $u_{RFG}^\mu$,
which is normal to the freeze out surface in the three dimensional space, and
differs from the Local Rest frame (LR) velocities of $f^*_{FO}$, (i.e.,
$u_L^\mu$ and $u_E^\mu$) we introduce here $\tilde{n}(\mu,T) = 8 \pi T^3
e^{\mu/T} (2\pi \hbar)^{-3} $, \ $a= {m\over T}$, \ so that $\hat{n}(\mu,T)=
\tilde{n} a^2 K_2(a)/2$ is the invariant
scalar density of the symmetric massless J\"uttner gas, $b=
a/\sqrt{1-v^2}=a\g$, \ $v \equiv v_\sigma = d\sigma_0 /
d\sigma_x=u_{RFG}^1|_{RFF}$, \ $A = (2+2b+b^2) e^{-b}$, \ and $$ {\cal K}_n
(z,w) \equiv \frac{2^{n} (n)!}{(2n)!}\ z^{-n} \!\!\int_w^\infty\!\!\!dx \
(x^2 - z^2)^{n-1/2} \ e^{-x} \ , $$ i.e. ${\cal K}_n (z,z) = K_n (z) $.  When
evaluating the limit we used the relation \linebreak
${\cal K}_n (a,b) 
\stackrel{\scriptstyle a,b=0}{\longrightarrow} 
K_n (a) \stackrel{\scriptstyle a=0}{\longrightarrow} (n-1)!\,2^{n-1} a^{-n}$.
This baryon current may then be Lorentz transformed into the Eckart Local
Rest (ELR) frame of the post FO matter, which moves with $u^\mu_E = N^\mu /
(N^\nu N_\nu)^{1/2}= \gamma_E (1,v_E,0,0)|_{RFG}$ in the RFG,\ \ or
alternatively into the Rest Frame of the Freeze out front (RFF), where
$d\hat{\sigma}_\mu = (0,1,0,0)|_{RFF}$ and the velocity of the RFG is $u^\mu_{RFG}
= \gamma_\sigma (1, v, 0, 0)|_{RFF}$.  Then the Eckart flow velocity of the
matter represented by the cut J\"uttner distribution viewed from the RFF is
$u^\mu_E = \gamma_c (1, v_c, 0,0)|_{RFF}$, where $v_c = (v + v_E)/(1+v v_E)$.

\item[c]
We are considering fast particles, so that $v_{rel} \sim v$. A formula
similar to (\ref{Px}) is also obtained for massless particles.  

\end{notes}

\vfill\eject
\end{document}